\documentclass[aip,reprint]{revtex4-2}
\usepackage[utf8]{inputenc}
\usepackage[T1]{fontenc}
\usepackage{epsfig}
\usepackage{float}
\usepackage{mathptmx}
\usepackage{dcolumn}
\usepackage{bm}
\usepackage{graphicx}
\usepackage{times}
\usepackage{float}
\usepackage{amssymb}
\usepackage{amsmath}
\usepackage{amsfonts}
\usepackage{gensymb}
\usepackage{enumitem} 
\usepackage{setspace}
\usepackage[dvipsnames]{xcolor}
\usepackage{titlesec}
\usepackage{upgreek}
\usepackage{mathtools}
\usepackage[sort&compress]{natbib}

\setcitestyle{open={},close={},numbers,comma,super}

\usepackage[pagebackref=false]{hyperref}
\hypersetup{
    colorlinks  =   true,
    linkcolor   =   BlueViolet,
    citecolor   =   BlueViolet,
    filecolor   =   BlueViolet,
    urlcolor    =   BlueViolet}
\usepackage[capitalise,nameinlink]{cleveref}
\usepackage{hypcap} % useless with aps/aip style, use \capstart for each figures!
\usepackage{titlesec}
\titleformat{\section}[hang]{\small\bfseries\sffamily}{\thesection.}{0.5em}{\MakeUppercase}
\titlespacing{\section}{0pc}{1.2pc}{0.3pc}
\titlespacing{\subsection}{0pc}{1pc}{0.2pc}

\makeatletter
\renewcommand*{\fnum@figure}{{\normalfont\bfseries \figurename~\thefigure}}
\renewcommand*{\@caption@fignum@sep}{\textbf{ : }}
\makeatother

\begin{document}
%\title{Molecular beam epitaxy of single-crystal niobium nitride ultra-thin films for superconducting nanowire single-photon detectors}

\title{Epitaxial niobium nitride superconducting nanowire single-photon detectors}

\author{Risheng Cheng}
\thanks{These authors contributed equally to this work.}
\affiliation{Department of Electrical Engineering, Yale University, New Haven, CT 06511, USA}

\author{John Wright}
\thanks{These authors contributed equally to this work.}
\affiliation{Department of Materials Science and Engineering, Cornell University, Ithaca, New York 14853, USA.}

\author{Huili G. Xing}
%\email{grace.xing@cornell.edu}
\affiliation{Department of Materials Science and Engineering, Cornell University, Ithaca, New York 14853, USA.}
\affiliation{School of Electrical and Computer Engineering, Cornell University, Ithaca, New York 14853, USA.}
\affiliation{Kavli Institute for Nanoscale Science, Cornell University, Ithaca, New York 14853, USA.}

\author{Debdeep Jena}
\email{djena@cornell.edu}
\affiliation{Department of Materials Science and Engineering, Cornell University, Ithaca, New York 14853, USA.}
\affiliation{School of Electrical and Computer Engineering, Cornell University, Ithaca, New York 14853, USA.}
\affiliation{Kavli Institute for Nanoscale Science, Cornell University, Ithaca, New York 14853, USA.}

\author{Hong X. Tang}
\email{hong.tang@yale.edu}
\affiliation{Department of Electrical Engineering, Yale University, New Haven, CT 06511, USA}
\date{\today}

%%%%%%%%%%% Abstract %%%%%%%%%%%%%%%%%%%%%%%%%%%%
\begin{abstract}
Superconducting nanowires used in single-photon detectors have  been realized on amorphous or poly-crystalline films. Here, we report the use of single-crystalline NbN thin films for superconducting nanowire single-photon detectors (SNSPDs).  Grown by molecular beam epitaxy (MBE) at high temperature on nearly lattice-matched AlN-on-sapphire substrates, the NbN films exhibit high degree of uniformity and homogeneity. Even with relatively thick films, the fabricated nanowire detectors show saturated internal efficiency at near-IR wavelengths, demonstrating the potential of MBE-grown NbN for realizing large arrays of on-chip SNSPDs and their integration with AlN-based $\chi^{(2)}$ quantum photonic circuits.
\end{abstract}
\maketitle

%%%%%%%%%%%%%%%%%%%%%%%%%%%%%%%%%%%%%%%%%
%%%%%%%%%%%%%%%%%%%%%%%%%%%%%%%%%%%%%%%%%
%%%%%%%%%%%%%%%%%%%%%%%%%%%%%%%%%%%%%%%%%
%%%%%%%%%%%%%%%%%%%%%%%%%%%%%%%%%%%%%%%%%
%%%%%%%%%%%%%%%%%%%%%%%%%%%%%%%%%%%%%%%%%

Superconducting nanowire single-photon detectors (SNSPDs)\cite{goltsman_2001_first_SNSPD,Hadfield_2009_SPD_review} have become an indispensable resource for a range of quantum and classical applications due to their high detection efficiency over a broad spectrum\cite{nist_2013_93p_efficiency,nist_2019_mid-ir_detector_spectroscopy,berggren_2012_mid-ir_detector,nist_2019_98p_snspd}, ultra-fast speed\cite{pernice_2018_2DPC,simit_2019_16_pixel_detector}, exceptional timing performance\cite{Delft_2018_10ps_jitter_detector,jpl_2020_3ps_jitter,delft_2020_platform_snspd}, and ultra-low dark count noise\cite{schuck_2013_mHz_dark_count,NTT_2015_ultimate_darkcounts}. Two categories of superconducting materials have so far been used for the fabrication of high-efficiency SNSPDs -- poly-crystalline nitride, and amorphous alloy superconductors. SNSPDs patterned with thin-film amorphous superconducting materials, such as WSi\cite{nist_2013_93p_efficiency,nist_2011_first_wsi_detector} and MoSi\cite{switzerland_2018_mosi_detector,goltsman_2014_mosi_detector,hadfield_2016_MoSi_waveguide_detector}, have exhibited excellent homogeneity over a large device area\cite{nist_2020_wsi_microwire,Charaev2020,nist_2019_kilopixel_snspd} due to the absence of grain boundaries. However, they require relatively lower operation temperature and have lower maximum counting rates, resulting from longer hot spot relaxation time in comparison with the SNSPDs made from nitride superconductors, such as NbN\cite{simit_2017_92p_nbn_detector,pernice_2015_waveguide_snspd,pernice_2012_waveguide_SNSPD,cheng_2019_snspd_ald} and NbTiN\cite{delft_2017_92p_nbn_detector,cheng_2017_multiple_SNAP,cheng_2016_self_aligned_detector,nict_2017_nbtin_snap,Machhadani2019}. On the other hand, Nb(Ti)N-based detectors have shown relatively superior timing performance, demonstrating <3\,ps jitter measured with a short straight nanowire\cite{jpl_2020_3ps_jitter} and <8\,ps with a large-area meandered nanowires\cite{EsmaeilZadeh2020}. Despite these advantages, the homogeneity of Nb(Ti)N-SNSPDs are ultimately limited by the poly-crystalline nature of the sputtered Nb(Ti)N films, which could lead to non-uniform distribution of critical currents in a large array of single-photon detectors required for future integrated quantum photonic circuits.      

\begin{figure}[h]
\capstart
\centering\includegraphics[width=1\linewidth]{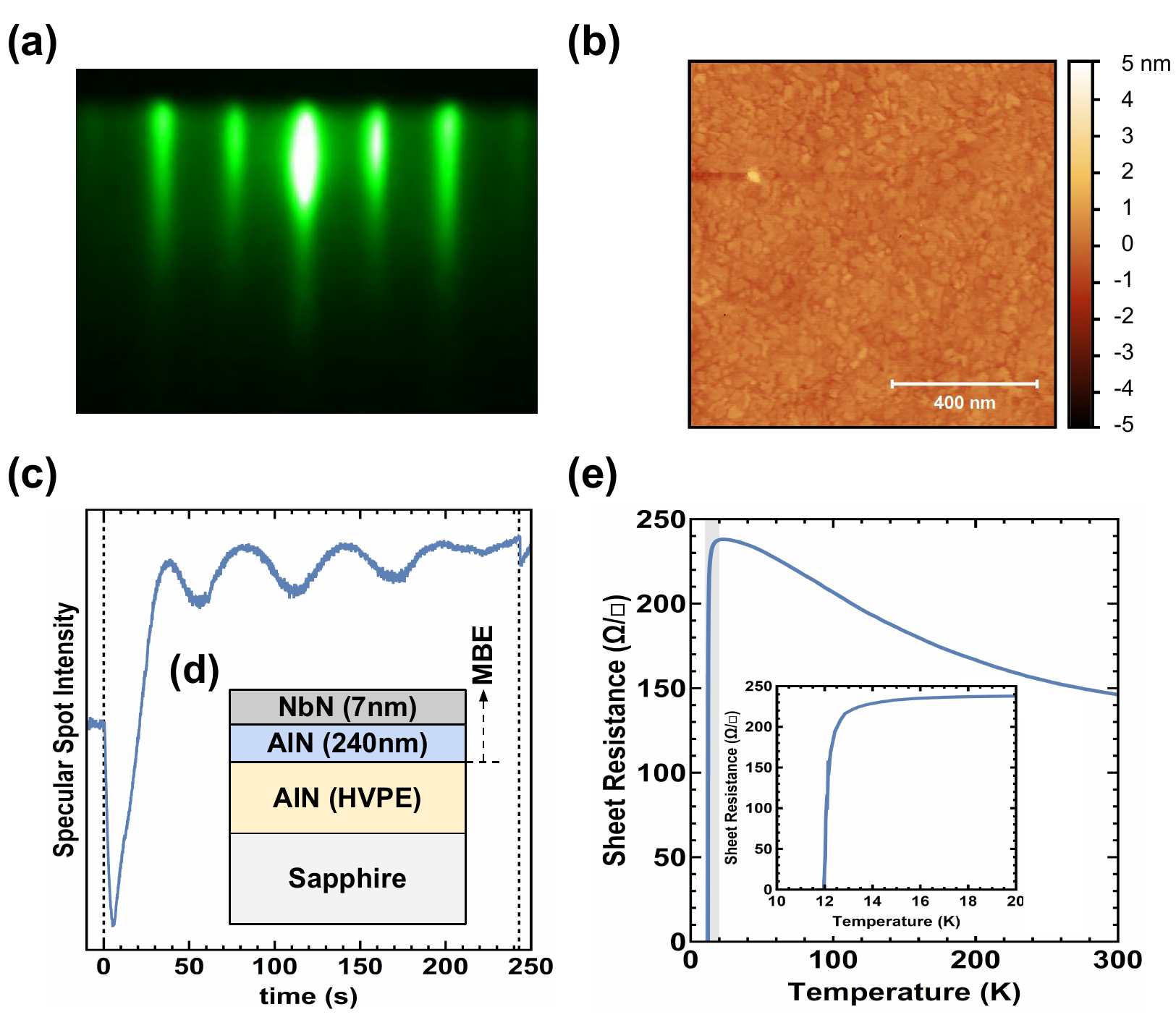}
\caption
{
(a) \textit{In-situ} RHEED pattern measured after the film growth demonstrating the epitaxial nature of the NbN film. The streakiness of the pattern evidences that the surface is effectively 2D.
(b) AFM surface height map of the NbN thin film exhibiting $R\textsubscript{rms}$=\,0.29\,nm.   
(c) RHEED intensity monitored throughout the NbN thin film growth. The exhibited oscillations of the specular spot brightness indicates the 2D layer-by-layer growth mode of the NbN. (d) Cross-sectional sketch of the thin film layer structure. (e) Measured sheet resistance of the NbN thin film versus temperature with the inset showing the $T_\mathrm{c}$ of 12.1\,K.
}
\label{fig:material}
\end{figure}

\begin{figure*}[!t]
\capstart
\centering\includegraphics[width=1\linewidth]{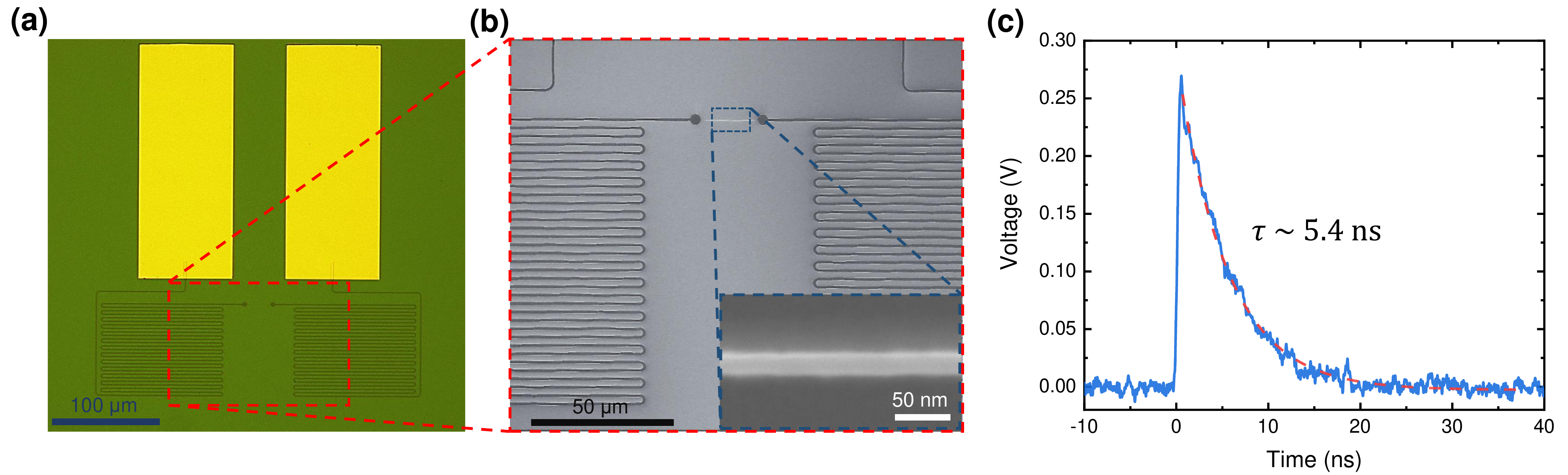}
\caption
{
(a) Optical micrograph image of the fabricated SNSPD device.
(b) Close-up scanning electron micrograph (SEM) image of the active straight nanowire and series inductor made of 1\,$\mu$m-wide meandered wire. The inset shows the further zoom-in view of the 20\,nm-wide straight nanowire.  
(c) Single-shot trace of output voltage pulse from the 20\,nm-wide detector measured by a 4\,GHz oscilloscope. The decay time constant ($\tau$) extracted from the exponential fitting (red dashed line) is 5.4\,ns. 
} 
\label{fig:device}
\end{figure*}

In this Letter, we demonstrate SNSPDs made from single-crystal NbN thin films grown by molecular beam epitaxy (MBE)\cite{Yan2018} on nearly lattice-matched AlN-on-sapphire substrates. This substrate platform is attractive for the integration of SNSPDs with several other elements of nitride-based photonic integrated circuits \cite{pernice_2018_waveguide_snspd_review,cheng_2019_broadband_spectrometer,italy_2019_amplitude_multiplex,berggren_2015_on_chip_detector,cheng_2020_perfect_absorber}. The epitaxial NbN films exhibit a high degree of thickness uniformity and structural perfection owing to the 2D layer-by-layer growth unique to MBE technique.
The fabricated device consisting of 20\,nm-wide and 7\,nm-thick nanowire shows saturated internal efficiency at the wavelength of 780\,nm and 1050\,nm, while further reduction in achievable thin film thickness holds promise for saturating the efficiency at longer wavelength with more relaxed wire width. We expect MBE-NbN on AlN-on-sapphire substrate shown here could provide a scalable material platform for realizing large array of on-chip SNSPDs and integration with nitride-based photonic circuits.

As illustrated in \cref{fig:material}, epitaxial NbN films are grown by radio-frequency plasma-assisted MBE on a commercial 2\,inch-diameter c-plane sapphire substrate with a 3\,$\mu$m-thick AlN film grown by hydride vapor phase epitaxy (HVPE). A 240\,nm-thick AlN film of Al-polar orientation is grown by MBE, followed by the growth of NbN as shown in \cref{fig:material}(d). During the growth of the films, the reactive nitrogen is generated using a radio-frequency plasma source fed by ultrahigh-purity N$_2$ gas, which is further purified by an in-line purifier. Aluminum (99.9999$\%$ purity) is supplied using a Knudsen effusion cell. The Nb flux is generated using an \textit{in situ} electron-beam evaporator source with 3N5-pure (excluding tantalum, Ta) Nb pellets in a tungsten hearth liner. The NbN films are grown at the temperature of 1100\,\degree C measured by a thermo-couple behind the substrate, at a growth rate of approximately 1.0\,nm/min.

The MBE film growth is monitored \textit{in situ} using a reflection high-energy electron diffraction (RHEED) system operated at 15\,kV voltage and 1.5\,A current. \Cref{fig:material}(a) shows sharp and streaky patterns formed by electron diffraction from the smooth surface of the NbN film, indicating the epitaxial nature of the single-crystal NbN film. As shown in \cref{fig:material}(c), the \textit{in situ} observation of oscillations of the RHEED intensity versus the growth time confirms that the NbN grows in a 2D layer-by-layer growth mode on the AlN surface. The thickness of the NbN film is 7.0\,nm,  measured by X-ray reflectivity (XRR) with a Rigaku SmartLab diffractometer using CuK$\alpha$1 radiation. \Cref{fig:material}(b) shows the morphology of the NbN film surface characterized employing tapping mode atomic force microscopy (AFM); the root-mean-square roughness ($R\textsubscript{rms}$) of the film surface is less than 0.3\,nm within a scan size of 1\,$\mu$m$\times$1$\mu$m. In addition, the crystal orientation of the NbN is determined using RHEED and X-ray diffraction (XRD), which indicates the cubic NbN grows with the $\{$1 1 1$\}$ crystal axis aligned to the c-axis of AlN.

\Cref{fig:material}(e) shows the temperature dependence of the sheet resistance of the MBE-NbN thin film with the inset showing a zoom-in view of the superconducting transition region. The transition temperature of the film is measured to be $T_\mathrm{c}=$12.1\,K, defined as the temperature where the resistance of the film drops to 50\% of the normal-state resistance measured at 20\,K. This value is among one of the highest results reported so far for NbN thin films of a few nanometer thickness. The high $T_\mathrm{c}$ value is also consistent with the significantly low resistivity of the film, which is calculated to be only $\sim$100$\,\mu\Omega\cdot\mathrm{cm}$ obtained by multiplying the thickness with the room-temperature sheet resistance.

We fabricate SNSPD devices by patterning the 7\,nm-thick MBE-NbN film. The nanowires are defined by the exposure of negative-tone 6\% hydrogen silsesquioxane (HSQ) resist using $100\,\mathrm{kV}$ electron-beam lithography (Raith EBPG\,5200) and the subsequent development in 25\% tetramethylammonium hydroxide (TMAH) for 2\,minutes at room temperature. The HSQ resist is spun at the speed of 4000\,rpm, resulting in an approximate thickness of 90\,nm. In a second electron-beam lithography step, contact electrodes are defined using double-layer polymethyl methacrylate (PMMA) positive-tone resist. After the development in the mixture of methyl isobutyl ketone (MIBK) and isopropyl alcohol (IPA), we liftoff electron-beam evaporated 10$\,$nm-thick Cr adhesion layer and 100$\,$nm-thick Au in acetone overnight to form the contact pads. The HSQ nanowire pattern is then transferred to the NbN layer in a timed reactive-ion etching (RIE) step employing CF\textsubscript{4} chemistry and 50\,W RF power. The HSQ resist is left on top of the NbN nanowires after fabrication, serving as a barrier to oxidation. 

For initial tests, we fabricate short-nanowire detectors with widths ranging from 20\,nm to 100\,nm for comparison of the internal efficiencies. As shown in \cref{fig:device}(a) and (b), the active detection parts of the devices are made of 20\,$\mu$m-long straight nanowires which are suitable for future waveguide integration. All the nanowires are serially connected to long 1\,$\mu$m-wide meandered wires to prevent the detector latching at high bias currents. The sheet resistance of the devices are measured to be around 180\,$\ohm$/sq, which slightly increase compared to the value measured on the bare film prior to fabrication.

In order to characterize the optical response of the fabricated detectors, the detector chip containing multiple devices is mounted on a 3-axis stack of Attocube stages inside a closed-cycle refrigerator and cooled down to 1.7\,K base temperature. Continuous wave (CW) laser light with varied wavelength is attenuated to the single-photon level and sent to the detector chip via a standard telecommunication fiber (SMF-28) installed in the refrigerator. The detectors are flood-illuminated by fixing the fiber tip far away from the surface of the detector chip. We control the Attocube stages to move the detector chip at low temperature and make the electrical contact between the RF probes and the gold pads of the detectors. The RF probes are connected to a semi-rigid coaxial cable installed in the refrigerator, while the room-temperature end of the cable is attached to a bias-tee (Mini-Circuits ZFBT-6GW+) to separate the DC bias current and RF output pulses for the detectors. The bias current is supplied by a programmable sourcemeter (Keithley 2401) in conjunction with a low-pass filter (1\,kHz cut-off frequency). The output pulses of the detectors are amplified by a low-noise RF amplifier (RF bay LNA-650) and sent to a 4\,GHz oscilloscope for the pulse observation or a pulse counter (PicoQuant PicoHarp 300) for the photon counting measurement. \Cref{fig:device}(c) shows a single-shot trace measurement of the output voltage pulse from the 20\,nm-wide detector. The decay time constant extracted from the exponential fitting (red dashed line) is 5.4\,ns, which translates into 24\,pH/sq sheet kinetic inductance of the NbN film, assuming 50\,$\ohm$ input impedance of the readout amplifier.   

\begin{figure}[!t]
\capstart
\centering\includegraphics[width=1\linewidth]{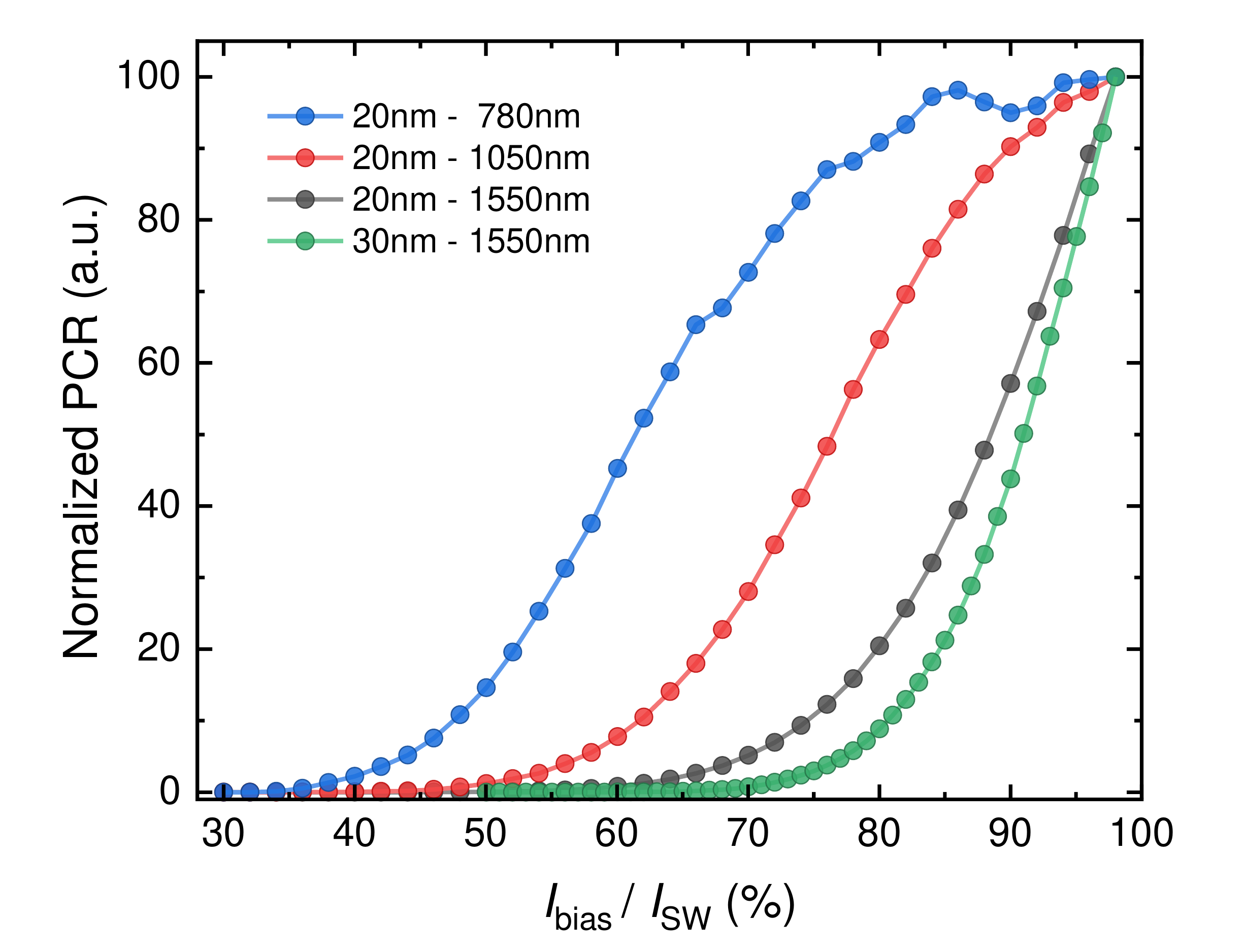}
\caption
{
Normalized photon counting rates (PCR) versus the relative bias current ($I_\mathrm{bias}/I_\mathrm{SW}$) measured with the 20\,nm-wide and 30\,nm-wide nanowire detectors for varying wavelength of photons. $I_\mathrm{SW}$ of the nanowires are measured to be 25.5\,$\mu$A and 38.8\,$\mu$A, respectively.
}
\label{fig:efficiency}
\end{figure}

\Cref{fig:efficiency} demonstrates the normalized photon counting rates (PCR) as a function of the relative bias current to the switching current ($I_\mathrm{bias}/I_\mathrm{SW}$) for 20\,nm-wide and 30\,nm-wide nanowire detectors. $I_\mathrm{SW}$ of the devices are measured to be 25.5\,$\mu$A and 38.8\,$\mu$A, respectively. As expected, detectors made of narrower nanowires with reduced $I_\mathrm{SW}$ show better saturated internal efficiencies at shorter wavelength. For 20\,nm-wide nanowire detector, we observe a broad saturation plateau at 780\,nm wavelength, while the efficiency is only nearly saturated at 1050\,nm wavelength. The minor fluctuation in the curve corresponding to 780\,nm wavelength is due to the polarization instability of the laser, since the photon absorption of the nanowire is significantly dependent on the polarization status of the incident photons. Neither the 20\,nm-wide nor the 30\,nm-wide nanowires show saturation behavior at 1550\,nm wavelength. We attribute the inefficiency of the detectors to the significantly low electrical resistivity of the MBE-NbN material. In comparison with the SNSPDs made from sputtered\cite{simit_2017_92p_nbn_detector} or atomic-layer-deposited (ALD)\cite{cheng_2019_snspd_ald,sayem_2019_ln_snspd} NbN film of comparable thickness, the nanowires shown in this work demonstrate approximately 3 times reduced sheet resistance as well as sheet kinetic inductance, and in the meantime, 3-5 times improved critical current density.  All of these results are in agreement with the 2-3 times lower resistivity of MBE-NbN compared to sputtered or ALD-NbN. Thus, we expect that by further reducing the MBE NbN film thickness down to 2-3\,nm, saturated efficiency can be obtained in longer wavelengths with relaxed nanowire widths up to 100\,nm. The growth of high crystalline quality NbN films of 3\,nm thick or less are achievable by MBE, although a method to protect such thin films from oxidation upon exposure to the ambient is necessary and under investigation. Future work will explore the suitability of such ultra-thin films for SNSPDs.    

In summary, we have demonstrated the first SNSPDs patterned from MBE-grown single-crystal NbN thin films on AlN. The 20\,nm-wide SNSPDs show saturated internal efficiency at the wavelength of 780 nm and 1050 nm. It is worth mentioning that the AlN-on-sapphire substrate, which the epitaxial growth of NbN relies on, is particularly attractive due to its potential of the on-chip integration of SNSPDs with versatile AlN nanophotonic circuits. The excellent optical functionalities of AlN, such as strong $\chi^{(2)}$/$\chi^{(3)}$ nonlinearity\cite{guo_2017_photon_pair} and large electro-optic effect\cite{fan_2018_eo_converter,xiong_2012_aln_review}, renders NbN on AlN-on-sapphire a very attractive material platform for realizing fully integrated quantum photonic circuits with the generation, routing, active manipulation and the final detection of single photons on a single chip.

%%%%%%%%%% Acknowledgments %%%%%%%%%%%%%%%%%%%%%
\section*{Acknowledgments}
This project is funded by Office of Naval Research grants (N00014-20-1-2126 and N00014-20-1-2176) monitored by Dr. Paul Maki.  D.J. acknowledges funding support from NSF RAISE TAQs Award No. 1839196 monitored by Dr. D. Dagenais. H.X.T. acknowledges funding support from DARPA DETECT program through an ARO grant (No: W911NF-16-2-0151), NSF EFRI grant (EFMA-1640959) and the Packard Foundation. The authors would like to thank Sean Rinehart, Kelly Woods, Dr. Yong Sun, and Dr. Michael Rooks at Yale University for their assistance provided in the device fabrication, and Dr. Scott Katzer and Dr. David Meyer at the Naval Research Laboratory for useful discussions. The epitaxial growth was performed at Cornell University, and material characterization used resources made available by the NSF CCMR MRSEC Award No. 1719875. The fabrication of the devices was done at the Yale School of Engineering \& Applied Science (SEAS) Cleanroom and the Yale Institute for Nanoscience and Quantum Engineering (YINQE).

%%%%%%%%%%%%%%%%%%%%%%%%%%%%%%%%%%%%%%%%%
%%%%%%%%%%%%%%%%%%%%%%%%%%%%%%%%%%%%%%%%%
%%%%%%%%%%%%%%%%%%%%%%%%%%%%%%%%%%%%%%%%%
%%%%%%%%%%%%%%%%%%%%%%%%%%%%%%%%%%%%%%%%%
%%%%%%%%%%%%%%%%%%%%%%%%%%%%%%%%%%%%%%%%%
\def\bibsection{\section*{References}}
\bibliographystyle{Risheng}
\bibliography{My_reference}
\end{document}